\documentclass[prl,aps,twocolumn, floatfix, superscriptaddress,showpacs]{revtex4-1}
\usepackage{amsmath,bm,graphicx}
\usepackage{amssymb}
\usepackage{amsfonts}

\newcommand{\pd}[2]{\frac{\partial #1}{\partial #2}}
\newcommand{\dd}[2]{\frac{d #1}{d #2}}

\begin{document}
\bibliographystyle{apsrev4-1}

\title{Mixed flux-equipartition solutions of a diffusion model of nonlinear cascades}

\author{Colm Connaughton}
\email{connaughtonc@gmail.com}
\affiliation{Warwick Centre for Complexity Science, University of Warwick, Coventry CV4 7AL, UK}
\affiliation{Mathematics Institute, University of Warwick, Coventry CV4 7AL, UK}
\author{Rachel McAdams}
\email{rm837@york.ac.uk}
\affiliation{Department of Physics, University of York, Heslington, York YO10 5DD, UK}

\date{\today}

\begin{abstract}
We present a parametric study of a nonlinear diffusion equation which generalises Leith's model of 
a turbulent cascade to an arbitrary cascade having a single conserved quantity. 
There are three stationary regimes depending on whether the Kolmogorov exponent 
is greater than, less than or equal to the equilibrium exponent. In the first regime,  the 
large scale spectrum scales with
the Kolmogorov exponent. In the second regime, the large scale spectrum scales with the 
equilibrium exponent so the system appears to be at equilibrium at large scales. Furthermore, 
in this equilibrium-like regime, the amplitude of the large-scale spectrum depends on 
the small scale cut-off. This is interpreted as an analogue of cascade nonlocality. In the 
third regime, the equilibrium spectrum acquires a logarithmic correction. An exact analysis of 
the self-similar, non-stationary 
problem shows that time-evolving cascades have direct analogues of these three regimes.
\end{abstract}

\maketitle


Cascades are often observed in the non-equilibrium statistical dynamics of 
interacting many-body systems in which microscopic interactions between degrees of 
freedom are conservative and sources and sinks of the conserved quantity are
widely separated. A famous example is the Richardson cascade in high Reynolds number 
hydrodynamic turbulence. There the nonlinear terms in the Navier-Stokes equation conserve the
fluid kinetic energy and energy injection (by stirring for example) and energy
dissipation (by viscosity) are widely separated in scale or wavenumber.
See \cite{FRI1995} and the references therein. In recognition of the historical importance
of this example, and for brevity, we shall always speak in this article of a cascade as 
the process whereby nonlinear interactions conservatively transport ``energy'', $E$, in 
``wavenumber'' space, $k$. We acknowledge that cascades occur in many other contexts in 
which the conserved quantity is not necessarily the energy and transport may occur in a
space other than the space of wave-numbers. Some examples include wave turbulence 
\cite{ZLF92}, cluster-aggregation \cite{LEY2003}, nonlinear diffusion \cite{VAZ2006} and 
Bose-Einstein condensation \cite{LLPR2001,CP04}.

The description of cascades based on the underlying dynamical equations typically
leads to Boltzmann-like kinetic equations obtained by moment closures
which may be phenomenological (as is common with hydrodynamic examples) or which may
be asymptotically exact (as with weak wave turbulence \cite{NR2011}). Transport in
such kinetic equations usually involves a nonlinear integral collision operator. This 
makes their detailed analysis difficult. One way around this difficulty, 
originally proposed by Leith \cite{LEI1967},  is to 
phenomenologically replace the integral collision operator with a more analytically 
tractable nonlinear differential 
operator in such a way as to preserve the scaling properties of the original problem and, in
particular, the scalings of the stationary Kolmogorov and equilibrium solutions. Such 
models are referred to as differential approximation models.
These phenomenological models allow the cascade dynamics to be qualitatively explored with 
relative ease. For that reason, they have become an active field of research in their own right.  The Leith model continues to
be of interest in turbulence \cite{CRW2010}, while similar models have been used 
to study two-dimensional turbulence \cite{LEI1968,LN2006}, wave turbulence \cite{HHAB1985,CNP03}, 
kelvin waves on vortex lines in a superfluid \cite{NAZ2006},  the Boltzmann equation for a 
hard-sphere gas \cite{POAN2011-arxiv} and optical turbulence \cite{DNPZ92}. They are often
used  as a heuristic way of establishing the direction of the Kolmogorov cascade
 \cite{DNPZ92,POAN2011-arxiv}, an issue which has caused controversy in 
some contexts. As we shall see below,
care must be taken to correctly interpret the predictions about the cascade direction made
using these models.

A disadvantage of differential approximation models is that, by construction, they model the 
nonlinear transport as a process which is local in scale. It is well known, however,  that
the integral character of the original collision integral may lead to cascade dynamics 
which are nonlocal in scale. That is to say the energy transfer through a given wavenumber is
dominated by interactions with smallest or largest wavenumber in the system rather than
with nearby wavenumbers. The phenomenological nature of differential approximation models
opens the door to the uncomfortable possibility of attempting to model a nonlocal cascade
with a local operator.

Consider the following generalisation of Leith's model:
\begin{equation}
\label{eq-genLeithModel}
\pd{E}{t} = -\pd{J}{k},
\end{equation}
where $J$ is the energy flux which is modeled as
\begin{equation}
\label{eq-flux}
J(k) = -k^{m\,x_K-x_T +1} E^{m-1} \pd{}{k}\left(k^{x_T}\,E\right).
\end{equation}
The sign of the flux is chosen such that the flux is positive when the energy flows to the 
right in $k$-space. This is easily verified by integrating Eq.~(\ref{eq-genLeithModel}) from
$0$ to $K$ (assuming that $J(0)=0$) and asking whether energy is entering or leaving the
interval $[0,K]$. Eq.~(\ref{eq-genLeithModel}) has three adjustable parameters, $m$, $x_K$ and
$x_T$. The parameter $m$, which we take to be greater than 1,  is the order of the nonlinear interaction responsible for the
transport of energy. The parameters $x_K$ and $x_T$ are, as we shall see below are the
exponents of the stationary Kolmogorov and thermodynamic equilibrium states, which we take to
be independent adjustable parameters in order to perform a parameteric study of the 
properties of the generalised Leith model. Note that the original Leith model is recovered
by setting $m=3/2$, $x_K$=$5/3$ and $x_T=-2$. Eq.~(\ref{eq-genLeithModel}) is appropriate
for modeling a system with a single conserved quantity, and thus a single cascade. We
should remark at this point that Eq.~(\ref{eq-genLeithModel}) is also of considerable
independent interest outside of its utility as a heuristic model of turbulent cascades.
For various values of the parameters, it appears as a model of flow in a porous medium
\cite{VAZ2006}, viscous gravity currents \cite{GM1990}, transport of density fluctuations
in a magnetised plasma \cite{NNR2005} and the spreading of surfactant on a liquid interface
\cite{JEN1995}.

The general stationary solution of Eq.~(\ref{eq-genLeithModel}) involves two constants which
we call $J$ and $T$:
\begin{equation}
\label{eq-stationarySolution}
E(k) = k^{-x_T}\left[T^m + \frac{J}{x_K-x_T}\,k^{(x_T-x_K)\,m} \right]^\frac{1}{m}.
\end{equation}
There are two stationary solutions which are pure power laws. The first, having
$J=0$ is 
\begin{equation}
\label{eq-thermodynamic}
E(k) = T\,k^{-x_T}.
\end{equation}
It corresponds to the thermodynamic equilibrium solution since the flux, Eq.~(\ref{eq-flux}), 
vanishes on this solution. For this reason, we refer to $T$ as the temperature even though,
strictly, the thermodynamic temperature is only defined at equilibrium. The second, having 
$T=0$, is
\begin{equation}
\label{eq-Kolmogorov}
E(k) = \left(\frac{J}{x_K-x_T}\right)^\frac{1}{m}\,k^{-x_K}.
\end{equation}
It corresponds to the Kolmogorov solution since the flux, Eq.~(\ref{eq-flux}), is constant and
equal to $J$ on this solution. 

From Eq.~(\ref{eq-Kolmogorov}) one can see that the flux, $J$,
carried by the Kolmogorov spectrum must be positive when $x_K>x_T$ corresponding to energy
transfer to the right in $k$-space and negative when $x_K<x_T$ corresponding to energy
transfer to the left in $k$-space. Energy transfer to the left is inconsistent with the
energy injection occuring at small $k$ and the energy dissipation occuring at small $k$ - that 
is the flux is in the ``wrong direction'' to connect the source and sink. The identification
of situations in which this happens is one of the popular applications of differential 
approximation models. The issue is not entirely theoretical and occurs in reality for
energy and particle cascades in the Boltzmann equation \cite{KKMN1975} and for the 
inverse cascade in two-dimensional optical turbulence \cite{DNPZ92}. It is generally agreed
that this means that the Kolmogorov spectrum is not physically realisable although there is
less consensus about what takes its place. We address this issue clearly and unambiguously
below, at least in the context of the generalised Leith model.

Let us return to the general stationary state, Eq.~(\ref{eq-stationarySolution}, which
has finite $J$ and $T$. There are three cases:

\noindent {\bf 1\ \  Kolmogorov-like regime, $\mathbf{x_K>x_T}$:} 
In this regime, we have a regular Kolmogorov cascade at large scales:
\begin{equation}
E(k) \sim \left(\frac{J}{x_K-x_T}\right)^\frac{1}{m}\,k^{-x_K} \hspace{1.0cm}\mbox{as $k\to 0$},
\end{equation}
which is thermalised at small scales:
\begin{equation}
E(k) \sim T\,k^{-x_T} \hspace{1.0cm}\mbox{as $k\to \infty$}.
\end{equation}
Such states are sometimes called ``warm'' cascades \cite{CN04} since they have a 
nonzero temperature parameter. They are relevant for the description of the statistical 
dynamics of the truncated Euler equations for example \cite{BBDB2005}.

\noindent {\bf 2\ \ Equilibrium-like regime, $\mathbf{x_K<x_T}$:} 
In this regime, the cascade has a completely different character. It appears
to be at equilibrium at large scales, despite carrying a constant flux:
\begin{equation}
E(k) \sim T\,k^{-x_T} \hspace{1.0cm}\mbox{as $k\to 0$}.
\end{equation}
Note that $J$ can be positive in this regime. Eq.~(\ref{eq-stationarySolution}) can 
therefore describe a cascade with the ``correct'' direction provided we allow a finite
value of $T$. The reason is that the term describing the flux is subleading. In contrast to 
the case $x_K>x_T$, the cascade has an intrinsic cut-off at which the spectrum vanishes given by
\begin{equation}
k_* = \left[\left(\frac{x_T-x_K}{J}\right)^\frac{1}{m} T\right]^\frac{1}{x_T-x_K}.
\end{equation}
It is, perhaps, more natural to consider the temperature, $T$, as a function
of the cut-off, $k_*$, which may be imposed for example by the dissipation scale. In this case, 
the stationary state can be written as
\begin{equation}
\label{eq-stationarySolution2}
E(k) = k^{-x_T}\left[\frac{J}{x_K-x_T}(k_*^{(x_T-x_K)\,m} - k^{(x_T-x_K)\,m} \right]^\frac{1}{m}.
\end{equation}
The amplitude of the spectrum at large scales depends on the small scale cut-off in
this regime. This is the analogue of non-locality for the differential approximation model.
Intriguingly, it was shown in \cite{CON2009} that the isotropic 3-wave kinetic equation is
always nonlocal when $x_K<x_T$. Thus not only does Eq.~(\ref{eq-stationarySolution2}) provide us
with an analogue of non-locality for Eq.~(\ref{eq-genLeithModel}) but it occurs for the
correct parameter regime. A relationship between the temperature and the small-scale cut-off
has recently been proposed and partially observed numerically in the context of the classical
Boltzmann equation \cite{POAN2011-arxiv}.

\noindent {\bf 3 Degenerate regime, $\mathbf{x_K=x_T}$:}
When the thermodynamic and Kolmogorov exponents coincide, the spectrum is:
\begin{equation}
\label{eq-ststionaryLog}
E(k) = (J\,m)^\frac{1}{m}\,k^{-x_T}\, \left(\log  \frac{k_*}{k}\right)^\frac{1}{m},
\end{equation}
so that the system appears to be at equilibrium with a logarithmic correction. This can be
seen by direct integration of the stationary version of Eq.~(\ref{eq-genLeithModel}).  It
is more informative to obtain this formula by Taylor expanding Eq.~(\ref{eq-stationarySolution2})
in small values of $x_K-x_T$ and taking the limit $x_K\to x_T$. By doing this, one sees clearly
that the logarithm is the remnant of the cut-off dependence of the equilibrium-like regime as 
one enters the cut-off independent Kolmogorov-like regime. 
While this case is a special point in the parameter 
space, it does occur in practice as for example in the direct cascade in 3-D NLS turbulence 
\cite{DNPZ92} and in elastic wave turbulence in a vibrating plate \cite{DJR2006}.

Let us now consider non-stationary cascades relevant to situations in which we do not
inject energy at large scales but rather consider the evolution of a lump of energy
which is initially concentrated at large scales. In this case, there is no stationary
spectrum and the evolution is described by a self-similar function of $k$ and $t$:
\begin{equation}
\label{eq-scaling}
E(k,t) = s(t)^a\,F(\xi)\hspace{1.0cm}\mbox{where $\xi=\frac{k}{s(t)}$},
\end{equation}
where $s(t)$ is a typical wavenumber which grows in time as the spectrum
spreads in $k$-space.  It is well-known (see for example \cite{LOT1982} and the references 
therein) that this self-similarity ansatz applied to nonlinear diffusion equations like 
Eq.~(\ref{eq-genLeithModel}) leads to weak solutions describing propagating fronts which are
 positive on an expanding compact interval, $[0,k_*(t)]$, and zero elsewhere. It is convenient 
therefore to take the characteristic scale, $s(t)$, to be the right boundary of the support of 
the solution corresponding to the front tip.  We show now that the same three regimes 
identified above for the stationary case have direct analogues in the non-stationary case.  
Substituting Eq.~(\ref{eq-scaling}) into Eq.~(\ref{eq-genLeithModel}) we obtain the scaling equations:
\begin{eqnarray}
\label{eq-scaling1} \dd{s}{t} &=& s^{m\,x_K + (m-1)\,a}\\
\label{eq-scaling2} a\,F - \xi\,^\dd{F}{\xi} &=& \dd{}{\xi}\left[\xi^{m\,x_K-x_T +1} F^{m-1} \dd{}{\xi}\left(\xi^{x_T}\,F\right) \right].
\end{eqnarray}
From conservation of energy, $\int_0^{k_*(t)} E(k,t)\,d k = 1$, Eq.~(\ref{eq-scaling}) leads
to $s^{a+1} \int_0^1 F(\xi)\, d \xi = 1$, from which we conclude that $a=-1$. To avoid the 
complications \cite{CN2010}  associated with self-similarity of the second kind, we shall assume from this
point on that $x_K<1$. Thus the results cited below are applicable to infinite capacity 
cascades only. When $a=-1$, the left hand side of Eq.~(\ref{eq-scaling2}) is an exact
differential and can be integrated explicitly to obtain the scaling function in
closed form (assuming that $1-mx_K+(m-1)x_T \neq 0$):
\begin{equation}
\label{eq-F}
F(\xi) = 
\left\{
\begin{array}{ll}
\xi^{-x_T}\left[\frac{ 1 - \xi^{(m-1)(x_T-x_{NS})}}{x_T-x_{NS}} \right]^\frac{1}{m-1}&\hspace{0.5cm}\mbox{if $0<\xi<1$}\\
0&\hspace{0.5cm}\mbox{otherwise}
\end{array}
\right.
\end{equation}
where we have introduced, for convenience, the nonstationary exponent
\begin{equation}
x_{NS} = \frac{m\,x_K-1}{m-1}.
\end{equation} 
This solution  is the analogue for the generalised Leith model of the front solutions of the 
porous medium equation originally obtained by Pattle \cite{PAT1959}. As before, examining
Eq.~(\ref{eq-F}) shows that there are 3 regimes:

\noindent {\bf 1 Nonstationary Kolmogorov-like regime, $\mathbf{x_{NS}>x_T}$:}
In this regime, we have a non-stationary cascade at large scales with the Kolmogorov
exponent, $x_K$, replaced by $x_{NS}$:
\begin{equation}
F(\xi) \sim \left(\frac{1}{x_{NS}-x_T}\right)^\frac{1}{m-1}\,\xi^{-x_{NS}} \hspace{1.0cm}\mbox{as $\xi\to 0$}.
\end{equation}

\noindent {\bf 2 Nonstationary equilibrium-like regime, $\mathbf{x_{NS}<x_T}$:}
In this regime the cascade appears to be at equilibrium at large scales, but with a
temperature which decays in time due to Eq.~(\ref{eq-scaling}):
\begin{equation}
F(\xi) \sim \left(\frac{1}{x_T-x_{NS}}\right)^\frac{1}{m-1}\,\xi^{-x_T} \hspace{1.0cm}\mbox{as $\xi\to 0$}.
\end{equation}

\noindent {\bf 3 Nonstationary degenerate regime, $\mathbf{x_K=x_T}$:}
When the nonstationary and equilibrium exponents coincide, we obtain the non-stationary
analogue of the logarithmic correction to the equilibrium scaling discussed above for the
stationary case:
\begin{equation}
\label{eq-logF}
F(\xi) = 
\left\{
\begin{array}{ll}
(m-1)^\frac{1}{m-1} \xi^{-x_T}\left[ \log \frac{1}{\xi}\right]^\frac{1}{m-1}&\hspace{0.5cm}\mbox{if $0<\xi<1$}\\
0&\hspace{0.5cm}\mbox{otherwise}
\end{array}
\right.
\end{equation}
Note  that, as in the stationary case, the flux is positive and to the right in all
three regimes. We could also consider nonstationary cascades with a source of energy, in
which case, the exponent $a$ would no longer be equal to $-1$ and we would lose the
exact differential on the left hand side of Eq.~(\ref{eq-scaling2}) which allowed us to
solve the problem explicitly. In this case, however, the asymptotics of the scaling function, 
$F(\xi)$, can be obtained using the phase plane methods developed in \cite{GM1990}. This
rather technical analysis will be presented elsewhere.
In order to connect the results presented here on the generalised Leith model back to the
integral collision operators which we purport to model, we remark that the nonstationary
regimes 1 and 3 have already been explored in considerable detail analytically and 
numerically in the context of the isotropic three-wave kinetic equation \cite{CK2010} 
and conform to the general behaviour outlined here.

To conclude, we have performed a complete parametric study of a nonlinear diffusion model
of a turbulent cascade with a single conserved quantity which generalises Leith's original
model of the energy cascade in 3 dimensional hydrodynamic turbulence.
Both stationary and non-stationary cascades can be described by simple analytic solutions
of the model. We showed that there are three regimes depending on whether the Kolmogorov
exponent is greater than, less than or equal to the equilibrium exponent.
In the Kolmogorov-like regime,
the equilbrium behaviour is a small correction to the finite flux spectrum at large scales.
Large scales are independent of the small scale cut-off. In the equilibrium-like regime, the
finite flux behaviour is a small correction to equilibrium spectrum at large scales.
The amplitude of the large scale spectrum is a diverging function of the small scale cut-off. This 
latter fact means that even differential approximation models can mimic some aspects of 
cascade nonlocality.  In the degenerate regime, both finite flux and equilibrium behaviours
are equally important leading to a logarithmic correction to the equilibrium spectrum. The
question of cascade direction is completely clear in this model. By allowing a finite $T$, the 
flux is always positive and in the ``correct'' direction.
The Kolmogorov-like and degenerate regimes are already well known but the equilibrium-like
regime has not been appreciated previously and should now be sought in kinetic equations
using the full collision integral. Finally, we remark that many of the interesting physical
applications of differential approximation models in which issues of cascade direction and
degeneracy of exponents arise have two conserved quantities. This considerably complicates
things because the corresponding differential equation is fourth order. We hope that
the comprehensive description of a single conservation law presented here will help
to clarify the issues arising in more complicated examples.
\section*{Acknowledgements}
We acknowledge useful discussions with D. Proment and S. Nazarenko.

%
\end{document}